# Lightweight Mediated Semi-Quantum Key Distribution Protocol with a Dishonest Third Party based on Bell States

Chia-Wei Tsai[1] and Chun-Wei Yang[2,*]

[1]Department of Computer Science and Information Engineering, National Taitung University, No. 369, Sec. 2, University Rd., Taitung 95092, Taiwan

[2]Center for General Education, China Medical University, No. 100, Sec. 1, Jingmao Rd., Beitun Dist., Taichung 406040, Taiwan

[2]Master Program for Digital Health Innovation, College of Humanities and Sciences, China Medical University, No. 100, Sec. 1, Jingmao Rd., Beitun Dist., Taichung 406040, Taiwan

[1]cwtsai@gm.nttu.edu.tw

[2,*]cwyang@mail.cmu.edu.tw

## Abstract

The mediated semi-quantum key distribution (MSQKD) protocol is an important research issue that lets two classical participants share secret keys securely between each other with the help of a third party (TP). However, in the existing MSQKD protocols, there are two improvable issues, namely (1) the classical participants must be equipped with expensive detectors to avoid Trojan horse attacks and (2) the trustworthiness level of TP must be honest. To the best of our knowledge, none of the existing MSQKD protocols can resolve both these issues. Therefore, this study takes Bell states as the quantum resource to propose a MSQKD protocol, in which the classical participants do not need a Trojan horse detector and the TP is dishonest. Furthermore, the proposed protocol is shown to be secure against well-known attacks and the classical participants only need two quantum capabilities. Therefore, in comparison to the existing MSQKD protocols, the proposed protocol is better practical.

## 1. Introduction

To establish a secure communication, any two participants must share a secret key. Therefore, the key distribution protocol is a fundamental part in cyber security research. In 1984, Bennet and Brassard [1] used the properties of quantum mechanics to propose the first quantum key distribution (QKD) protocol — BB84 [1]. Further, some studies [2-4] proved that the BB84 protocol is unconditionally secure. In the classical cryptography, only one-time pad can conform to the unconditionally secure ciphers. Following the BB84 protocol, various QKD protocols [5-15] have been proposed. However, these QKD protocols assume that the protocol participants have complete quantum capabilities, implying that the participants can generate any type of quanta (single photons or entanglement states), store these qubits in quantum memory, and measure the qubits using any basis, among others. Most of these quantum capabilities are expensive and they are difficult to implement at present. To improve the practicality of the QKD protocol, Boyer et al. [16, 17] defined the semi-quantum concept and proposed the first semi-quantum key distribution (SQKD) protocol that consists of two types of participants: the quantum participant and the classical participant. The quantum participant has complete quantum capabilities, whereas the classical participant only owns limited quantum capabilities. After the semi-quantum environment was proposed, various kinds of semi-quantum protocols have been proposed for different security issues, some of which are SQKD for different situations [18-26], semi-quantum communication [27-33], semi-quantum secret sharing [34-37], semi-quantum private comparison [38, 39], and semi-quantum information splitting [40], among others[41-43]. According to the existing semi-quantum protocols [16-43], this study summarizes the semi-quantum environments and the quantum capabilities of the classical participants in **Table 1**. It is worth noting that the unitary operations for a single qubit have been taken as candidates [26, 41-43] that the classical participants can utilize (i.e., the unitary operation based environment in **Table1**) except for the four operators proposed by Boyer et al. [16, 17]. This is because the related technology of the unitary operations for a single qubit has developed rapidly and there are many quantum computers (e.g., the IBM Q system, Amazon Braket and so no) that provide mature quantum unitary operations.

**Table 1**. Summary of semi-quantum environment

| Environment | Capabilities of classical user |
| --- | --- |
| Measure-Resend Environment | ● generating Z-basis qubits<br>● Z-basis measurement<br>● reflecting photons without disturbance |
| Randomization-Based Environment | ● Z-basis measurement<br>● reordering photons using different delay lines<br>● reflecting photons without disturbance |
| Measurement-Free Environment | ● generating Z-basis qubits<br>● reordering photons using different delay lines<br>● reflecting photons without disturbance |
| Unitary Operation Based Environment | ● Z-basis measurement<br>● generating Z-basis qubits<br>● performing unitary operations |

Although SQKD protocols are more practical than the QKD protocol, existing SQKD protocols only focus on allowing two classical participants to share a secret key. However, letting two classical participants distribute secure keys is another interesting research topic. To handle this issue, Krawec [21] proposed the first mediated semi-quantum key distribution (MSQKD) protocol. In this protocol, two classical participants can distribute a secret key with the help of a quantum third party (TP). Here, the trustworthiness of a TP can be categorized into the four levels summarized in **Table 2,** based on [44]. The trustworthiness of TP in Krawec's protocol belongs to a dishonest TP.

**Table 2.** Trustworthiness levels of TP

| Trustworthiness Level | Definition |
|---|---|
| Honest TP | The TP has to follow the procedure of the protocol honestly and the participants can completely trust it. Therefore, the participants can share their secret information with the TP. However, the assumption of a trustworthy TP may be impractical. |
| Semi-honest TP | The TP has to execute the protocol loyally, but it may try to obtain the participants' secret information passively using the records of all intermediate transmissions and computations by the participants. |
| Almost dishonest TP | To extract the participants' secret information, the TP may perform any possible attacks except collaborating with other participants. This assumption is only suitable for some applications such as a quantum private comparison protocol. |
| Untrusted/dishonest TP | The TP may perform any possible attacks. |

After Krawec's MSQKD protocol, Liu et al. [24] used the entanglement swapping of Bell states [45] to improve the efficiency of the MSQKD protocol. Lin et al. [25] used single photons to design the MSQKD protocol to make it even more practical. Recently, Francesco Massa et. al. [46] proposed an efficient MSQKD protocol, in which the classical participants only have the two capabilities including detecting and reflecting the qubits. In the above-mentioned MSQKD protocol, the TP and classical participants adopt the two-way quantum communication to distribute key information. In comparison to the one-way quantum communication, the two-way quantum communication results in two issues. (1) The classical participants need additional quantum devices (e.g., the photon number splitter or the optical wavelength filter) to screen out the Trojan photons. Equipping these devices may violate the original intention of the semi-quantum environment (i.e., reducing the quantum capabilities of the classical participants). (2) Because the transmission time of qubits is more than doubled, the qubits in the two-way quantum communication exhibit easier decoherence than the one-way communication. Taking IBM Q Melbourne as an example, the average decoherence time for 16 qubits, T1 (for maintaining energy) and T2 (for maintaining phase), are 65.30 and 22.70 μs, respectively. This indicates that if the transmission time is greater than T1 or T2, the original information in the qubits cannot be obtained. Therefore, the TP and the participants must spend more to maintain the qubits in a two-way quantum communication. Tsai et al. [26] proposed a lightweight MSQKD protocol without the abovementioned issues; however, the trustworthiness of the TP is assumed to be honest, which may be impractical.

In this study, we refer the concept of [5] to design a mediated semi-quantum key distribution protocol with a dishonest TP. In the proposed protocol, the TP takes Bell states $\left|\Phi^{+}\right\rangle$ as quantum resources to assist the two classical participants in distributing the secret key but the TP cannot obtain any information about this secret key even if it performs any possible attack. In contrast, the classical participants only need two quantum properties including (1) Z-basis measurement and (2) performing Hadamard operator. The two capabilities have been practiced in quantum computers [47] or optical experiment implementation [48-53]; that is to say, Hadamard operation and Z-basis measurement devices have feasibilities in real implementation. Therefore, the proposed protocol maintains the lightweight property in terms of the quantum capabilities of the classical participant. Moreover, the one-way quantum communication strategy is adopted to design the protocol, and thus the proposed protocol is immune to Trojan horse attacks, implying that the classical participants do not equip any Trojan Horse detector.

The rest of this paper is organized as follows. Section 2 introduces the quantum properties used in the proposed protocol and the proposed lightweight mediated semi-quantum key distribution (LMSQKD) protocol. Section 3 presents the security analyses of the proposed LMSQKD protocol and then provides comparisons between the state-of-the-art MSQKD protocols in Section 4. Finally, the conclusions are presented in Section 5.

## 2. Proposed LMSQKD Protocol

In this section, the assumptions, quantum capability limitations of the classical participants, and the quantum properties used in the proposed protocol are described, and a lightweight mediated semi-quantum key distribution protocol is proposed.

In this study, we assume that two classical participants, Alice and Bob, want to share the secret key with the help of a TP, where the TP is dishonest (i.e., TP may perform any possible attack to compromise the distributed key). There are the one-way quantum channels between the TP and each classical participant (i.e., Alice and Bob). The classical channel between Alice and Bob is assumed to be authenticated. This study assumes that a classical participant has two quantum capabilities including (1) measuring the qubit using Z-basis $\{|0\rangle, |1\rangle\}$ and (2) performing Hadamard operator H, where H is defined as follows:

$$H = \frac{1}{\sqrt{2}}\left(|0\rangle\langle 0| + |1\rangle\langle 0| + |0\rangle\langle 1| - |1\rangle\langle 1|\right) = \frac{1}{\sqrt{2}}\begin{bmatrix} 1 & 1 \\ 1 & -1 \end{bmatrix} \quad \text{Eq. (1).}$$

However, the TP needs to generate the Bell states $|\Phi^{+}\rangle$, which are defined as follows:

$$|\Phi^{+}\rangle = \frac{1}{\sqrt{2}}\left(|00\rangle + |11\rangle\right) \quad \text{Eq. (2).}$$

The related assumptions and limitations in this study are summarized in **Table 3**.

**Table 3**. Summary of Assumptions and Limitations

| Assumption and Limitation | Description |
| --- | --- |
| Capacities of classical users | (1) Performing H operation<br>(2) Measuring qubit using Z-basis |
| Capabilities of TP | Generating Bell states |
| Trueness of TP | Dishonest |
| Quantum channel | (1) TP and Alice have an one-way quantum channel, TP→Alice<br>(2) TP and Bob have an one-way quantum channel, TP→Bob |
| Classical channel | (1) Alice ⇆ Bob is an authenticated classical channel.<br>(2) TP ⇆ Alice and TP ⇆ Bob are the authenticated classical channels. |

To enable Alice and Bob to share the secret key with the help of the TP, this study uses a quantum property: the relationship between Bell states and Hadamard operator. Here, Alice and Bob randomly decide to implement the identity operator I (i.e., do nothing) or Hadamard operator H on one of the two qubits (i.e., Alice implements the unitary operator on the first qubit of the Bell states and Bob implements the unitary operator on the second qubit). Then, they measure the qubits using Z-basis. The relationships between their implemented operators and measurement results are summarized in the following table, where $mr_A$ and $mr_B$ denote Alice's and Bob's measurement results, respectively.

**Table 4**. Relationship between Bell states and Hadamard operations

| **Initial state** | **Alice's operation** | **Bob's operation** | **Qubit State** | **Relationship of measurement result** |
|---|---|---|---|---|
| $\lvert\Phi^+\rangle_{AB} = \frac{1}{\sqrt{2}}(\lvert 00\rangle + \lvert 11\rangle)_{AB}$ | I | I | $\frac{1}{\sqrt{2}}(\lvert 00\rangle + \lvert 11\rangle)_{AB}$ | $mr_A = mr_B$ |
| | I | H | $\frac{1}{\sqrt{2}}(\lvert 00\rangle + \lvert 01\rangle + \lvert 10\rangle - \lvert 11\rangle)_{AB}$ | Uncertain |
| | H | I | $\frac{1}{\sqrt{2}}(\lvert 00\rangle + \lvert 01\rangle + \lvert 10\rangle - \lvert 11\rangle)_{AB}$ | Uncertain |
| | H | H | $\frac{1}{\sqrt{2}}(\lvert 00\rangle + \lvert 11\rangle)_{AB}$ | $mr_A = mr_B$ |

According to the above-mentioned relationship, we can determine the two cases as follows.

**Case 1**: Alice and Bob can use their measurement results as the raw key bits or checking bits when they perform the same operations.

**Case 2**: When they use different operations, they will discard the measurement results owing to the uncertain measurement results (i.e., Alice cannot use her measurement results to infer Bob's).

It should be noted that the measurement results are pure-random values in accordance with the property of measurements in Bell states, implying that Alice and Bob will share a one-time pad key (pure-random key) using this quantum property. That is, the participants do not spend the extra cost for generating the pure-random key.

This study assumes that Alice and Bob want to distribute an n-bit secret key. The steps of the proposed LMSQKD protocol are described as follows (also shown in Figure 1).

**Step 1.** TP generates the Bell state $\lvert\Phi^+\rangle$ and then, sends the first and second qubits of the Bell state to Alice and Bob, respectively.

**Step 2.** After receiving the qubit, Alice (Bob) performs H or I operation on the qubit with the probabilities of $P_a$ ($P_b$) or $1-P_a$ ($1-P_b$), respectively. And then, they measure the qubits using Z-basis to obtain the measurement results $mr_A^i$ and $mr_B^i$, where i indicates the i-th time measurement result.

TP, Alice, and Bob repeat **Step 1** and **Step 2** 4n times.

**Step 3.** According to **Table 4**, Alice and Bob discard the useless measurement results and choose the enough remaining measurement results as the checking bits (e.g., using 50% measurement results) to perform **Public Discussion** [1] to detect outsider or insider attackers using the authenticated classical channel. If the error rate is more than the pre-defined threshold (the threshold will be evaluated in the Section 3.1.2), they will terminate the protocol and restart from the beginning; otherwise, they will continue the protocol.

**Step 4.** Alice and Bob use the remaining measurement results in **Step 3** and the relationship shown in **Table 4** to obtain the raw key bits, and then perform privacy amplification [54, 55] on the raw key bits to obtain the distributed secret key. Generally, Alice and Bob will obtain $n/4$-bit secret key if $P_a = 0.5$ and $P_b = 0.5$ in Step 2, and they use the half corresponding measurement results as the checking bits in **Step 3**.

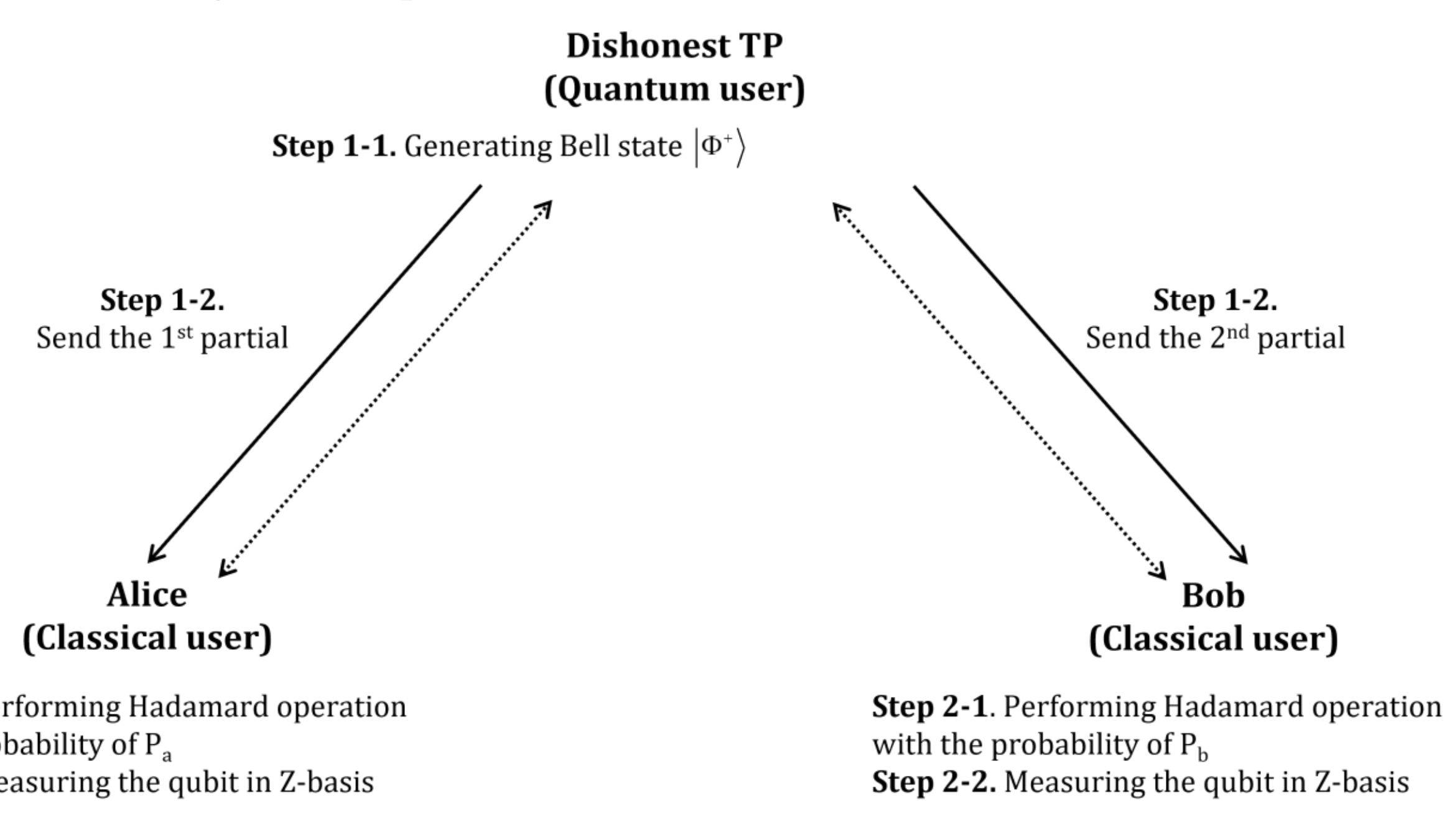

**Figure 1**. Processes of the proposed LMSQKD protocol

## 3. Security Analysis

In this section, we analyze the security of the proposed LMSQKD protocol. In terms of security analysis, the collective attack is a very important class of attacks, and the assumption of attacker's power in the collective attack is more powerful than the individual attack (e.g., the intercept-and-resend attack) [56, 57]. Thus, in this section, a complete collective attack analysis is given first, followed by the analysis of TP's fake photons attack and, the Trojan horse attack.

### 3.1. Collective Attack

For the collective attack, there are two types of analyses. In the first type of analysis, we need to prove that the attacker will disturb the original quantum system if the attacker wants to obtain useful information (i.e., robustness defined by [16, 17]). In the second type of analysis, the amount of information that the attacker can obtain is analyzed. In this study, we want to use the first type of analysis to prove that the collective attack will occur the rise in quantum bit error rate (QBER) and the participants abort the protocol when QBER is more than a preset threshold t, and then the information-theoretic security method is adopted to evaluate the proposed protocol's the key rate bound which will be used to formulate the pre-defined threshold t.

#### 3.1.1 Robustness analysis

For robustness analysis, we prove that an attacker cannot perform a collective attack to obtain any information regarding the raw key without being detected by the participants in the proposed MSQKD protocol. It should be noted that a dishonest TP has more advantages than an outside attacker. Therefore, we consider the TP as an attacker to discuss the security of the proposed protocol.

Before analyzing this attack, we first define the collective attack as follows:

(1) TP can insert its ancillary qubits in each quantum system transmitted on the quantum channel and then measure the ancillary qubits to obtain Alice's or Bob's secret key bit.

(2) Each quantum system sent between the users is attacked by the TP independently using the same strategy.

(3) The TP can keep the ancillary qubits until any later time, implying that it can measure the ancillary qubits after obtaining some information originating from this attack.

Therefore, a dishonest TP will perform a unitary operation $U_E$ to entangle the initial quantum system with its prepared ancillary qubits $\mathrm{E}=\{|\mathrm{E}_0\rangle,|\mathrm{E}_1\rangle,\ldots,|\mathrm{E}_{n-1}\rangle\}$ and measure them later to obtain useful information from the proposed protocol by performing the collective attack. $U_E$ must comply with the theorems of quantum mechanics, and thus it is defined as follows.

$$U_E\left(|\Phi^+\rangle\otimes|\mathrm{E}_\mathrm{i}\rangle\right)=a_0|00\rangle|e_0\rangle+a_1|01\rangle|e_1\rangle+a_2|10\rangle|e_2\rangle+a_3|11\rangle|e_3\rangle \qquad \text{Eq. (3),}$$

where $|\mathrm{E}_\mathrm{i}\rangle$ denotes the initial state of TP's ancillary qubit; $|e_0\rangle$, $|e_1\rangle$, $|e_2\rangle$, and $|e_3\rangle$ are four states that can be distinguished by the TP (i.e., the four states are orthogonal to each other); and $|a_0|^2+|a_1|^2+|a_2|^2+|a_3|^2=1$. Because Alice and Bob discard the measurement results when they implement different operators, we only consider the following two situations: (1) Alice and Bob both implement the I operator and (2) they both implement the H operator.

For the first situation, the quantum system can be given as follows.

$$a_0|00\rangle|e_0\rangle+a_1|01\rangle|e_1\rangle+a_2|10\rangle|e_2\rangle+a_3|11\rangle|e_3\rangle \quad \text{Eq. (4).}$$

Because $|e_0\rangle$, $|e_1\rangle$, $|e_2\rangle$, and $|e_3\rangle$ are the four states that can be distinguished by TP, it can infer Alice's and Bob's measurement results using the ancillary qubits. By linearity, the quantum system of the second situation can be shown as follows.

$$\frac{1}{2}\begin{bmatrix} |00\rangle\otimes\left(a_0|e_0\rangle+a_1|e_1\rangle+a_2|e_2\rangle+a_3|e_3\rangle\right) \\ +|01\rangle\otimes\left(a_0|e_0\rangle-a_1|e_1\rangle+a_2|e_2\rangle-a_3|e_3\rangle\right) \\ +|10\rangle\otimes\left(a_0|e_0\rangle+a_1|e_1\rangle-a_2|e_2\rangle-a_3|e_3\rangle\right) \\ +|11\rangle\otimes\left(a_0|e_0\rangle-a_1|e_1\rangle-a_2|e_2\rangle+a_3|e_3\rangle\right) \end{bmatrix} \quad \text{Eq. (5).}$$

Here, TP can also distinguish the four states $a_0|e_0\rangle+a_1|e_1\rangle+a_2|e_2\rangle+a_3|e_3\rangle$, $a_0|e_0\rangle-a_1|e_1\rangle+a_2|e_2\rangle-a_3|e_3\rangle$, $a_0|e_0\rangle+a_1|e_1\rangle-a_2|e_2\rangle-a_3|e_3\rangle$, and $a_0|e_0\rangle-a_1|e_1\rangle-a_2|e_2\rangle+a_3|e_3\rangle$ because these states are still orthogonal to each other, implying that TP can also obtain Alice's and Bob's measurement results in the second situation. In the proposed protocol, Alice and Bob take the public discussion to check their measurement result in **Step 3.** Thus, TP must adjust $U_E$ to avoid the participants' check. If TP adjusts $U_E$ for the first situation, it will set $a_1$ and $a_2$ as 0 to avoid the classical participants' public discussion. However, according to this setting, the quantum system of the second situation will be given as follows.

$$\frac{1}{2}\begin{bmatrix} |00\rangle\otimes\left(a_0|e_0\rangle+a_3|e_3\rangle\right) \\ +|01\rangle\otimes\left(a_0|e_0\rangle-a_3|e_3\rangle\right) \\ +|10\rangle\otimes\left(a_0|e_0\rangle-a_3|e_3\rangle\right) \\ +|11\rangle\otimes\left(a_0|e_0\rangle+a_3|e_3\rangle\right) \end{bmatrix} \quad \text{Eq. (6).}$$

To pass through the classical participants' public discussion, TP also must set $a_0|e_0\rangle-a_3|e_3\rangle=\vec{0}$, which signifies $a_0|e_0\rangle=a_3|e_3\rangle$, implying that TP cannot obtain any information about Alice's and Bob's measurement results in both situations. In contrast, if TP adjusts $U_E$ for the second situation, it will set $a_0|e_0\rangle-a_1|e_1\rangle+a_2|e_2\rangle-a_3|e_3\rangle=a_0|e_0\rangle+a_1|e_1\rangle-a_2|e_2\rangle-a_3|e_3\rangle=\vec{0}$, which signifies $a_0|e_0\rangle=a_3|e_3\rangle$. Then, TP sets $a_1$ and $a_2$ as 0 for the first situation. After the abovementioned setting, the quantum systems can be given as follows:

$$a_0|00\rangle|e_0\rangle+a_3|11\rangle|e_3\rangle \quad \text{Eq. (7),}$$

$$\frac{1}{2}\begin{bmatrix} |00\rangle \otimes \left(a_0|e_0\rangle + a_3|e_3\rangle\right) \\ |11\rangle \otimes \left(a_0|e_0\rangle + a_3|e_3\rangle\right) \end{bmatrix} \text{ Eq. (8).}$$

According to the analyses above, although the TP can determine a collective attack path $U_E$, and Alice and Bob cannot detect the attack, the TP cannot obtain any information regarding the raw key because $a_0|e_0\rangle = a_3|e_3\rangle$. In contrast, if the TP wishes to obtain useful information regarding the classical participants' raw key, the TP cannot use $U_E$ to execute a collective attack, so the TP's attack will induce a detectable disturbance that increases the QBER. This gives Alice and Bob a nonzero probability of detecting the TP's attack.

### 3.1.2 Key rate bound evaluation

To analyze the bound of secret-key rate in the semi-quantum key distribution protocol, Krawec [58, 59] have proposed the applicable key-rate proof manner for two-way quantum communication. However, because the qubit transmission is one-way in the proposed protocol, we only use the security analysis method proposed in [60] to evaluate the lower bound of the secret-key rates, in which the lower bound of the secret-key rate is proposed as followed:

$$r \geq \sup_{U \leftarrow A} \inf_{\sigma_{AB} \in \Gamma_{QBER}} \left(S\left(U \mid E\right) - H\left(U \mid B\right)\right) \text{ Eq. (9).}$$

In this formula, $r := \lim_{n\to\infty}\left(\ell^{(n)}/n\right)$, $S\left(U \mid E\right)$ denotes the von Neumann entropy of U (i.e., raw key bits of Alice) conditioned on an attacker's probe system (here, we assume TP as the attacker). $H\left(U \mid B\right)$ is Shannon entropy of U conditioned on Bob's measurement results B. In the proposed protocol, Alice and Bob have the two measurement modes, that is, (1) Model 1: measuring the qubit by Z-basis immediately, (2) Mode2: performing Hadamard operator on the qubit and then measuring it by Z-basis (note that this measurement is equivalent to X-basis measurement). Therefore, the key rate bound evaluation method of the propose protocol is similar to BB84's evaluation method proposed in [57]. After TP's collective attack, we set the quantum sates as

$$|\Psi\rangle_{ABE} := \sum_{i=1}^{4} \sqrt{\lambda_i}\, |\Phi_i\rangle_{AB} \otimes |v_i\rangle_E \text{ Eq. (10),}$$

where $|\Phi_1\rangle_{AB} \ldots |\Phi_4\rangle_{AB}$ denote the four Bell states in Alice and Bob's joint system and $|v_1\rangle_E \ldots |v_4\rangle_E$ are some mutually orthogonal states in TP's probe system. Assuming the quantum bit error rate (QBER) is Q, we can get $\lambda_3 + \lambda_4 = Q$ (with respect to Mode 1) and $\lambda_2 + \lambda_4 = Q$ (with respect to Mode 2). Normalizing $\lambda_1 + \lambda_2 + \lambda_3 + \lambda_4 = 1$, $\lambda_3 + \lambda_4 = Q$ and $\lambda_2 + \lambda_4 = Q$, we get $\lambda_1 = 1 + \lambda_4 - 2Q$, $\lambda_2 = \lambda_3 = Q - \lambda_4$ and $\lambda_4 \in [0, Q]$. Because the evaluation methods and results are the same in the two measurement modes, we only describe the analysis processes of the first measurement mode (measuring the qubit by Z-basis) in this section.

Let $\left|\theta^{a,b}\right\rangle$ denotes the state of TP's probe system, where a and b denote Alice's and Bob's measurement results, respectively. $\left|\theta^{a,b}\right\rangle$ could be the four kind of states (with respect to Mode 1) shown as followed:

$$\begin{aligned}
\left|\theta^{0,0}\right\rangle &= \frac{1}{\sqrt{2}}\left(\sqrt{\lambda_1}\left|v_1\right\rangle_E + \sqrt{\lambda_2}\left|v_2\right\rangle_E\right) \\
\left|\theta^{0,1}\right\rangle &= \frac{1}{\sqrt{2}}\left(\sqrt{\lambda_3}\left|v_3\right\rangle_E + \sqrt{\lambda_4}\left|v_4\right\rangle_E\right) \\
\left|\theta^{1,0}\right\rangle &= \frac{1}{\sqrt{2}}\left(\sqrt{\lambda_3}\left|v_3\right\rangle_E - \sqrt{\lambda_4}\left|v_4\right\rangle_E\right) \\
\left|\theta^{1,1}\right\rangle &= \frac{1}{\sqrt{2}}\left(\sqrt{\lambda_1}\left|v_1\right\rangle_E - \sqrt{\lambda_2}\left|v_2\right\rangle_E\right)
\end{aligned} \quad \text{Eq. (11).}$$

According to Eq. (11), we can get the density operator of the TP probe system as followed:

$$\sigma_{TP}^{a} = \begin{pmatrix} \lambda_1 & \pm\sqrt{\lambda_1\lambda_2} & 0 & 0 \\ \pm\sqrt{\lambda_1\lambda_2} & \lambda_2 & 0 & 0 \\ 0 & 0 & \lambda_3 & \pm\sqrt{\lambda_3\lambda_4} \\ 0 & 0 & \pm\sqrt{\lambda_3\lambda_4} & \lambda_4 \end{pmatrix} \quad \text{Eq. (12),}$$

where $\pm$ is a plus if $a=0$ and a minus if $a=1$. Therefore, according to Eq. (9) and (12), we can get

$$S(U\,|\,E) - H(\mathrm{U}\,|\,\mathrm{B}) = \left(S(E\,|\,U) - S(E)\right) - \left(H(\mathrm{B}\,|\,\mathrm{U}) - H(\mathrm{B})\right) \quad \text{Eq. (13),}$$

with $S(E\,|\,U) = \frac{1}{2}S\left(\sigma_{TP}^{0}\right) + \frac{1}{2}S\left(\sigma_{TP}^{1}\right)$, $S(E) = S\left(\frac{1}{2}\sigma_{TP}^{0} + \frac{1}{2}\sigma_{TP}^{1}\right)$, $H(B\,|\,U) = h(Q)$ and $H(B) = 1$, where $h()$ is the binary entropy function. We use the Eq. (13) to evaluate the secret key rate of the proposed protocol and obtain that the secret key rate is a positive rate if $Q \le 0.11$. **Figure 2** shows the secret key rates under the different QBER values.

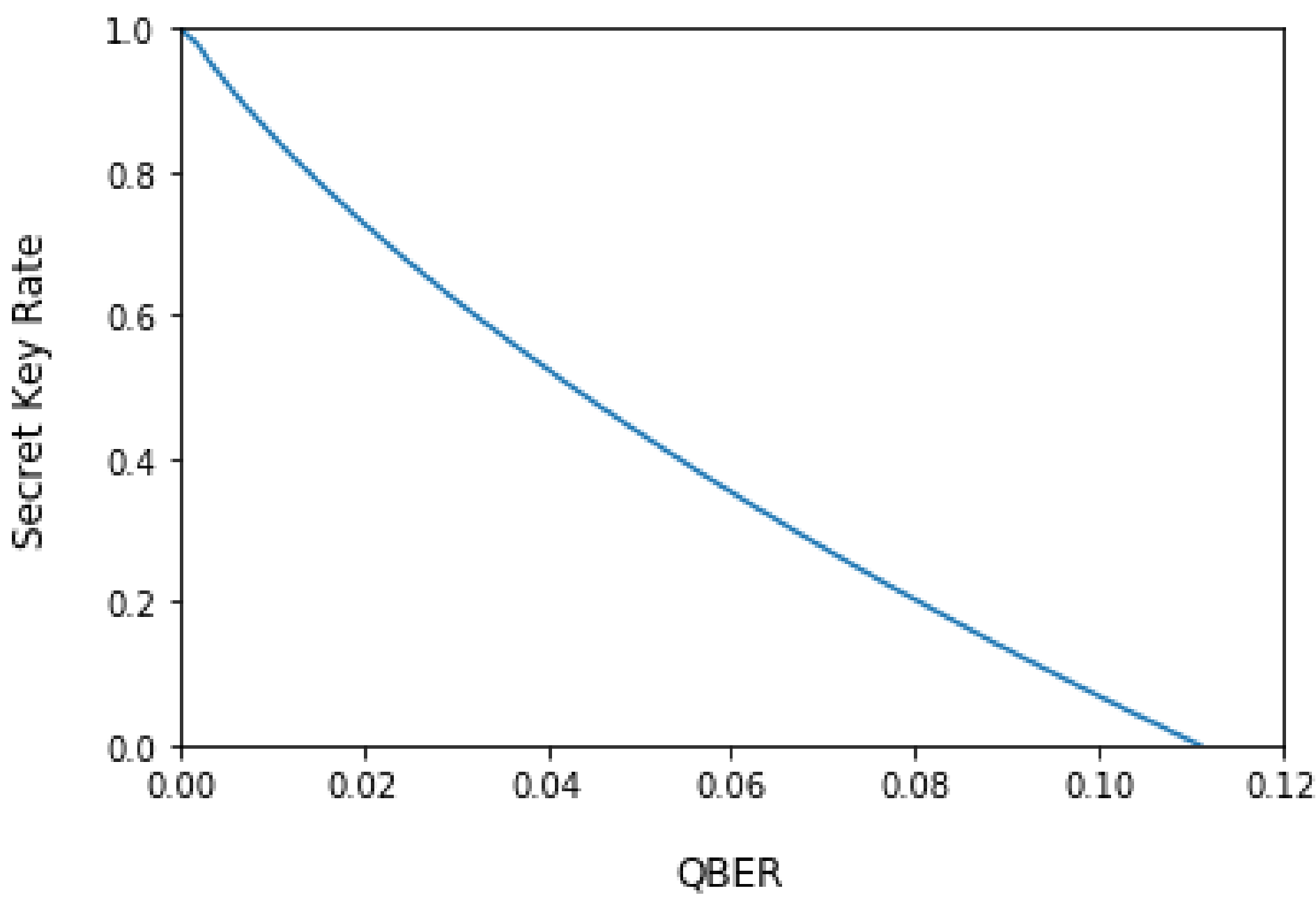

**Figure 2**. The secret key rate under the different QBER values

### 3.2 TP's fake photon attack

In the proposed protocol, Alice and Bob use the measurement results of Z-basis to be the secret keys. Therefore, in addition to inserting the ancillary qubits in each quantum system transmitted on the quantum channel, TP can also use the other quantum system instead of the Bell state to manipulate the classical participants' measurement results for stealing their secret keys.

The TP can take a single photon pair using Z-basis instead of Bell states, where the states of a single photon pair are generated depending on the original Bell state $|\Phi^+\rangle$. For example, TP will generate the photon pair with the same state (e.g., $\{|0\rangle,|0\rangle\}$ or $\{|1\rangle,|1\rangle\}$), and then TP sends the first and second qubits of the single photon pair to Alice and Bob in **Step 1** of the proposed protocol, respectively. Because the photon pairs are generated by TP, it can determine the classical participants' measurement results to manipulate the raw key bits. Unfortunately, TP's attack can be detected by Alice and Bob in **Step 3** because TP has no information about the operators implemented by Alice and Bob in **Step 2**. This implies that when both Alice and Bob implement the H operator on the single photon pair, their measurement results may violate the relationship shown in **Table 4** with the probability of $1/2$. Taking an example to explain this situation, TP generates $\{|0\rangle,|0\rangle\}$ instead of Bell state $|\Phi^+\rangle$ and then, the classical participants both implement the H operator. Here, their measurement results may be one of the four possibilities $\{|0\rangle,|0\rangle\}$, $\{|0\rangle,|1\rangle\}$, $\{|1\rangle,|0\rangle\}$, and $\{|1\rangle,|1\rangle\}$. If the measurement result is $\{|0\rangle,|1\rangle\}$ or $\{|1\rangle,|0\rangle\}$, Alice and Bob can detect TP's attack by Public Discussion. Although TP generates single photon pairs using X-basis ($|+\rangle=\frac{1}{\sqrt{2}}(|0\rangle+|1\rangle),|-\rangle=\frac{1}{\sqrt{2}}(|0\rangle-|1\rangle)$)

to avoid the abovementioned detection, its attack can still be determined by Alice and Bob when they both implement the I operator in **Step 2**. The detection probability for the fake photon attack is $1-\left(\frac{1}{2}\times\frac{1}{2}\right)^{m}=1-\left(\frac{1}{4}\right)^{m}$, where m denotes Alice and Bob using m measurement results to do Public Discussion. It implies that if m is large enough, the detection probability will approach 1, that is, the proposed protocol is robust against fake photon attack.

### 3.3 Trojan horse attack

Regarding implementation-dependent attacks, Trojan horse attacks [61, 62] are common. This paper discusses two types of Trojan horse attacks. The first is a delayed photon Trojan horse attack. In this attack, an attacker intercepts a qubit transmitted to a participant and then inserts a probing photon in the qubit with a delay time that is shorter than the time window. In this method, a participant cannot detect the fake photon because it does not register on their detector. After a participant performs the corresponding operation and returns the qubit, the attacker intercepts the qubit again and separates the probing photon. In this case, the attacker can obtain full information regarding a participant's operation by measuring the probing photon. The second attack is an invisible photon Trojan horse attack. The main strategy of this attack is to insert an invisible photon in each qubit sent to the participant. Because the participant's detector cannot detect this photon and performs a unitary operation on the qubit (the invisible photon also performs the same operation simultaneously), the attacker can steal information regarding the participant's operations in a manner similar to the delay photon Trojan horse attack.

In the attack methods mentioned above, the attacker can only extract information regarding participant operations when they retrieve the Trojan horse photons. A two-way communication protocol gives the attacker a chance to retrieve Trojan horse photons. Therefore, a protocol is only vulnerable to Trojan horse attacks if it adopts two-way communication. In contrast, in a one-way communication protocol, the attacker has no chance to retrieve the Trojan horse photons because no qubit is returned by the participant. In other words, the protocol will be robust to Trojan horse attacks if it is a one-way communication protocol.

In the proposed MSQKD protocol, the qubit quantum transmission strategy only operates in one direction, meaning qubits are only sent from the TP to the classical participants. Although the attacker can insert probing photons into the original qubits, they cannot extract any information regarding the participants' secret keys because the probing photons cannot be retrieved. Therefore, the proposed protocol is immune to Trojan horse attacks. Therefore, the classical participants do not need to be equipped with expensive devices (such as photon number splitters and optical wavelength filter devices) to mitigate Trojan horse attacks.

## 4. Performance Comparison

This section presents a comparison between the existing state-of-the-art MSQKD including Krawec's [21], Liu et al.'s [24], Lin et al.'s [25], and Tsai et al.'s [26] protocols. The comparison includes the semi-quantum environment, classical participant's quantum capabilities, quantum resources, trustworthiness level of TP, quantum efficiency, time of maintaining qubit to avoid decoherence, and whether a classical participant needs to be equipped with Trojan Horse detectors.

In terms of the semi-quantum environment and classical participant's quantum capabilities, both Krawec's and Lin et al.'s protocols use the measure-resend environment to ensure that the classical participants do not store the qubits. Although the classical participants do not need the quantum measurement devices, they must store the qubits for a period time to reorder them. The protocols proposed by Tsai et al. and this study only let the classical users be equipped with two quantum devices to ensure that this environment is more lightweight than the unitary operation-based environment. This type of environment is called lightweight unitary operation-based environment. For quantum resources, besides the protocols proposed by Lin et al. and Tsai et al., the TP needs to use the Bell states as quantum resources to assist Alice and Bob in distributing the secret keys. It should be noted that the TPs in these protocols are dishonest, except for Tsai et al.'s protocol.

Furthermore, we compared the proposed MSQKD to other protocols in terms of qubit efficiency, which is defined by the following equation [63-65].

$$QE = \frac{b_s}{q_t} \quad \text{Eq. (13)},$$

where $b_s$ denotes the number of bits of the shared session key and $q_t$ denotes the number of total qubits used in the protocol. In our protocol, TP generates n Bell states to let Alice and Bob share $n/4$-bit secret key if they set $P_a = 0.5$ and $P_b = 0.5$ in **Step 2**, and use the half corresponding measurement results as the checking bits in Step 4. Therefore, the qubit efficiency of our protocol is $\frac{n}{4}/2n = 1/8$, which is better than the others besides Liu et al.'s protocol.

This study assumes the time taken by the TP to transmit a qubit to each classical participant to be t. It should be noted that the qubits reflected by each participant to the TP are the same. Because the protocols proposed by Krawec, Lin et al., and Liu et al. use the two-way quantum communication, TP and the participants spent at least 2t time to maintain the qubits to avoid decoherence. However, Liu et al.'s protocol requires the qubits to be reordered. Therefore, the time taken to maintain the qubits increases to r, to reorder the qubits, and thus the maintain time should increase r which is the time of reordering the qubit sequence. In contrast, the protocols proposed by Tsai et al. and this study only spend time t to maintain the qubits because they adopted the one-way quantum communication. **Figure 3** shows the time taken to maintain the qubits for two-way and one-way quantum communications. Excluding the protocols proposed by Tsai et al. and this study, the classical participants need to be equipped with Trojan horse attack detectors.

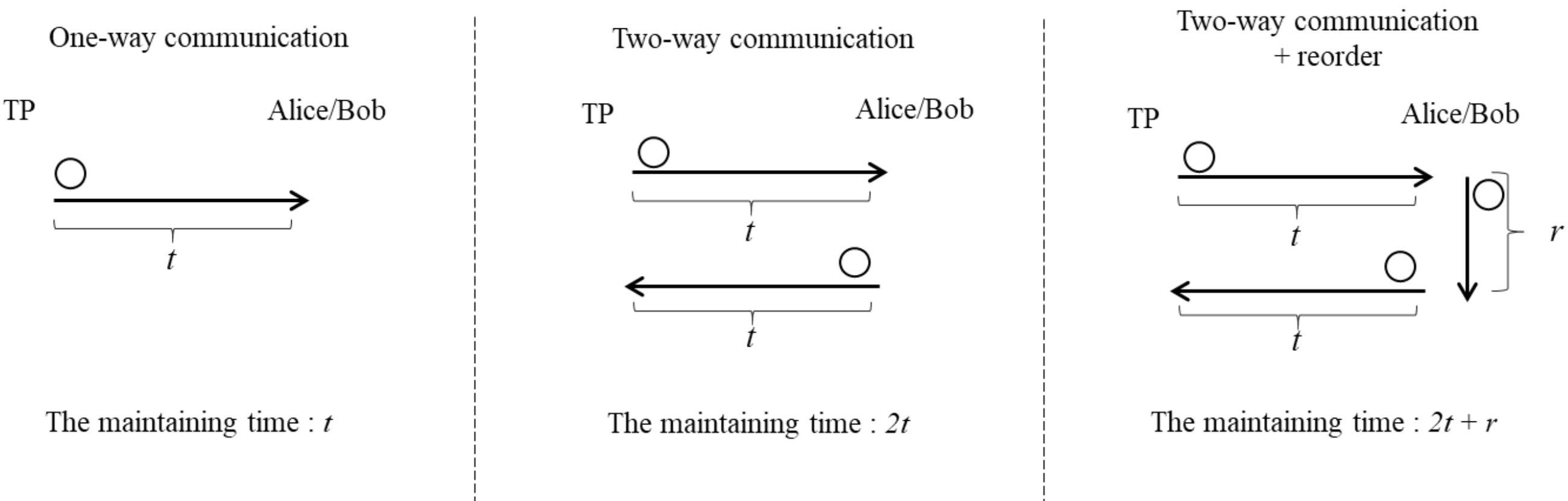

**Figure 3.** Time of maintaining qubits

All comparisons are summarized in **Table 5**. According to the above-mentioned comparison issues, our protocol is better than other MSQKD protocols in terms of practical implementation.

**Table 5.** Comparison to the existing MSQKD protocols

| | **Krawec's [21]** | **Liu et al.'s [24]** | **Lin et al.'s [25]** | **Tsai et al.'s [26]** | **Our protocol** |
|---|---|---|---|---|---|
| Semi-quantum environment | Measure-Resend | Measurement-Free | Measure-Resend | Lightweight Unitary Operation | Lightweight Unitary Operation |
| Quantum capability of classical participant | (1) Generation<br>(2) Measurement<br>(3) Reflection | (1) Generation<br>(2) Reflection<br>(3) Reorder | (1) Generation<br>(2) Measurement<br>(3) Reflection | (1) Measurement<br>(2) Operation | (1) Measurement<br>(2) Operation |
| Quantum resources | (1) Single photon<br>(2) Bell state | (1) Single photon<br>(2) Bell state | Single photon | Single photon | Bell state |
| Trustworthiness level of TP | Dishonest | Dishonest | Dishonest | Honest | Dishonest |
| Qubit efficiency | 1/24 | 1/8 | 1/24 | 1/32 | 1/8 |
| Time for maintaining qubits | 2t | 2t+r | 2t | t | t |
| Equipping with detectors or not | Yes | Yes | Yes | No | No |

## 5. Conclusions

To make the mediated key distribution protocol more practical, this study referred the concept of [5] to propose a lightweight mediated semi-quantum key distribution protocol to enable the sharing of secret keys between two classical participants with the help of a dishonest TP, in which the classical participants only need to be equipped with two quantum devices. The proposed protocol adopts one-way quantum communication to reduce the time of preventing the qubits from decoherence and avoid the use of Trojan horse detectors. The security analysis and performance comparison are presented to demonstrate that the proposed protocol is secure and efficient. The proposed protocol only let two classical participants to share secret keys, and thus how to let multiple participants can share the group keys is our future research.

## Acknowledgments

We would like to thank the anonymous reviewers and the editor for their very valuable comments, which greatly enhanced the clarity of this paper. This research was partially supported by the Ministry of Science and Technology, Taiwan, R.O.C. (Grant Nos. MOST 110-2221-E-039-004, MOST 110-2221-E-143-003, MOST 110-2221-E-259-001, and MOST 110-2221-E-143-004), and China Medical University, Taiwan (Grant No. CMU110-MF-121).